\documentclass{article}

\usepackage{arxiv}
\usepackage{graphicx}
\usepackage{dcolumn}
\usepackage{bm}
\usepackage{placeins}
\usepackage{subcaption}
\usepackage{stackengine}
\usepackage{amsmath}
\usepackage{multirow}
\usepackage{array}
\usepackage{amsfonts}
\usepackage{ifpdf}

\title{Spin Photonics in 3D Whispering Gallery Mode Resonators}

\author{Farhad Khosravi$^{1,2}$, Cristian L. Cortes$^{2}$, and Zubin Jacob$^{1,2}$\\
$^{1}$Department of Electrical and Computer Engineering, University of Alberta,\\ Edmonton, Alberta T6G 1H9, Canada \\
$^{2}$Birck Nanotechnology Center, School of Electrical and Computer Engineering,\\ Purdue University, West Lafayette, IN 47906, USA } 

\begin{document}
\maketitle

\begin{abstract}
Whispering gallery modes are known for having orbital angular momentum, however the interplay of local spin density, orbital angular momentum, and the near-field interaction with quantum emitters is much less explored. Here, we study the spin-orbital interaction of a circularly polarized dipole with the whispering gallery modes (WGMs) of a spherical resonator. Using an exact dyadic Green's function approach, we show that the near-field interaction between the photonic spin of a circularly polarized dipole and the local spin density of whispering gallery modes gives rise to unidirectional behaviour where modes with either positive or negative orbital angular momentum are excited. We show that this is a manifestation of spin-momentum locking using the whispering gallery modes of the spherical resonator. We discuss the requirements for possible experimental demonstrations using Zeeman transitions in cold atoms or quantum dots, and outline potential applications of these previously overlooked properties. Our work firmly establishes local spin density, momentum and decay as a universal right-handed electromagnetic triplet for evanescent waves.
\end{abstract}

\section{Introduction}

Spin-momentum locking is the source of unidirectional chiral phenomena in both electronic and photonic systems\cite{lin2013polarization,bliokh2014extraordinary,barik2018topological,li2014electrical,rodriguez2013near,le2015nanophotonic,petersen2014chiral,aiello2009transverse}. In topological insulators, spin-polarized edge modes have a spin direction that is dependent on the propagation direction of the modes \cite{li2014electrical,li2016electrical}. In photonics, the near-field interaction between a circularly polarized emitter and a metal interface gives rise to the unidirectional propagation of surface plasmon polaritons \cite{rodriguez2013near,le2015nanophotonic,sollner2015deterministic,young2015polarization,huang2013helicity}. This unidirectionality has also been observed in optical fibers. For example, in the experimental demonstration \cite{petersen2014chiral}, the near-field coupling between Zeeman transitions in a cold caesium atom and the $HE_{11}$ modes of the fibre was shown to give rise to the unidirectional propagation of light within the fiber. Optical resonators with broken symmetry also show unidirectional behaviour when coupled to a waveguide.
Examples of broken symmetries include the deformation of a resonator \cite{redding2012local}, broken time-reversal symmetries \cite{aoki2006observation}, as well as the mechanical spinning of a resonator \cite{maayani2018flying}.

\begin{figure}[b!]
        \centering
        \begin{subfigure}[t]{0.5\textwidth}
        \includegraphics[width=\linewidth]{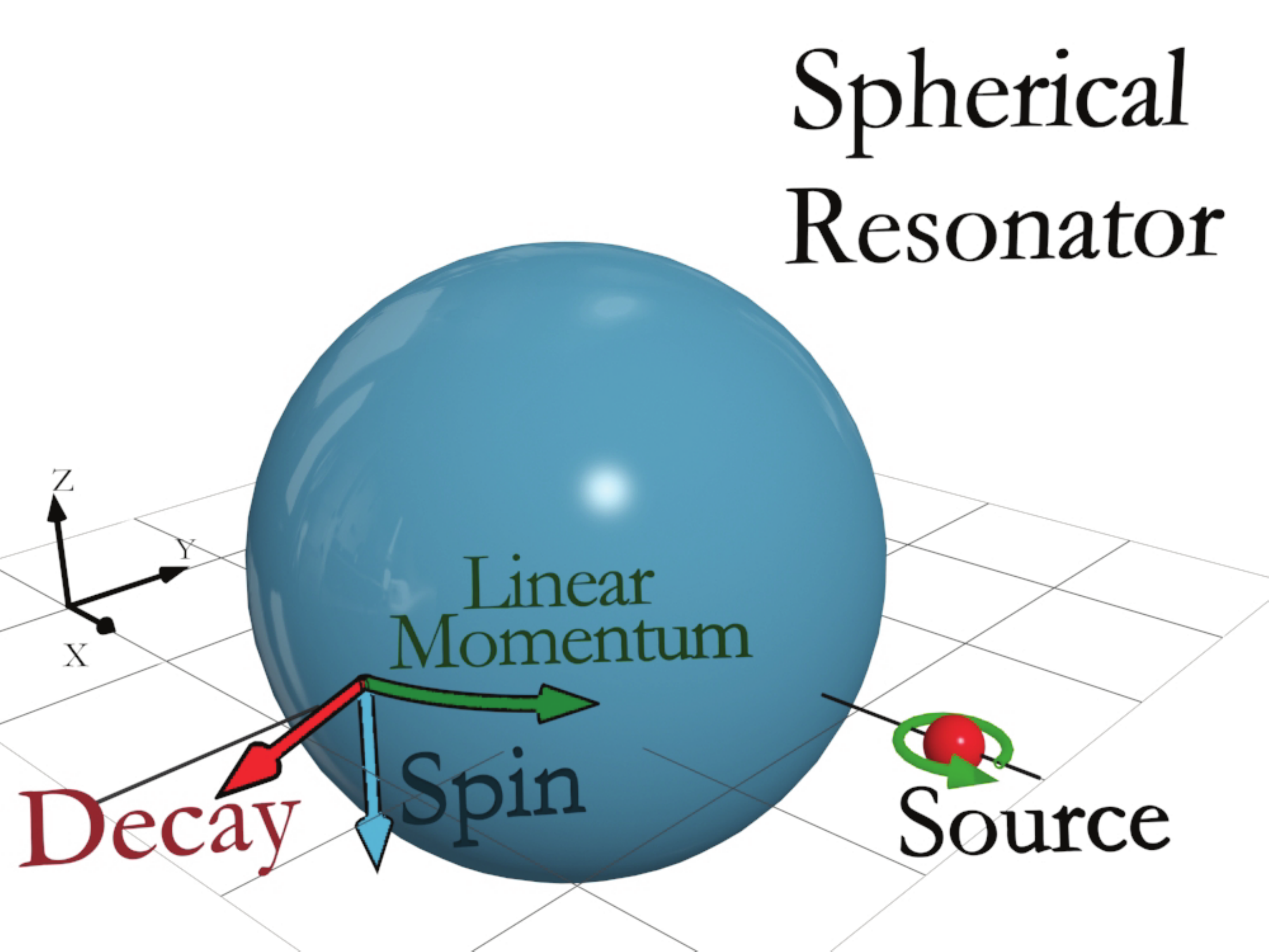}
        \caption{}
        \label{Fig:Schematic}
        \end{subfigure}
        \begin{subfigure}[t]{0.43\textwidth}
        \includegraphics[width=\linewidth]{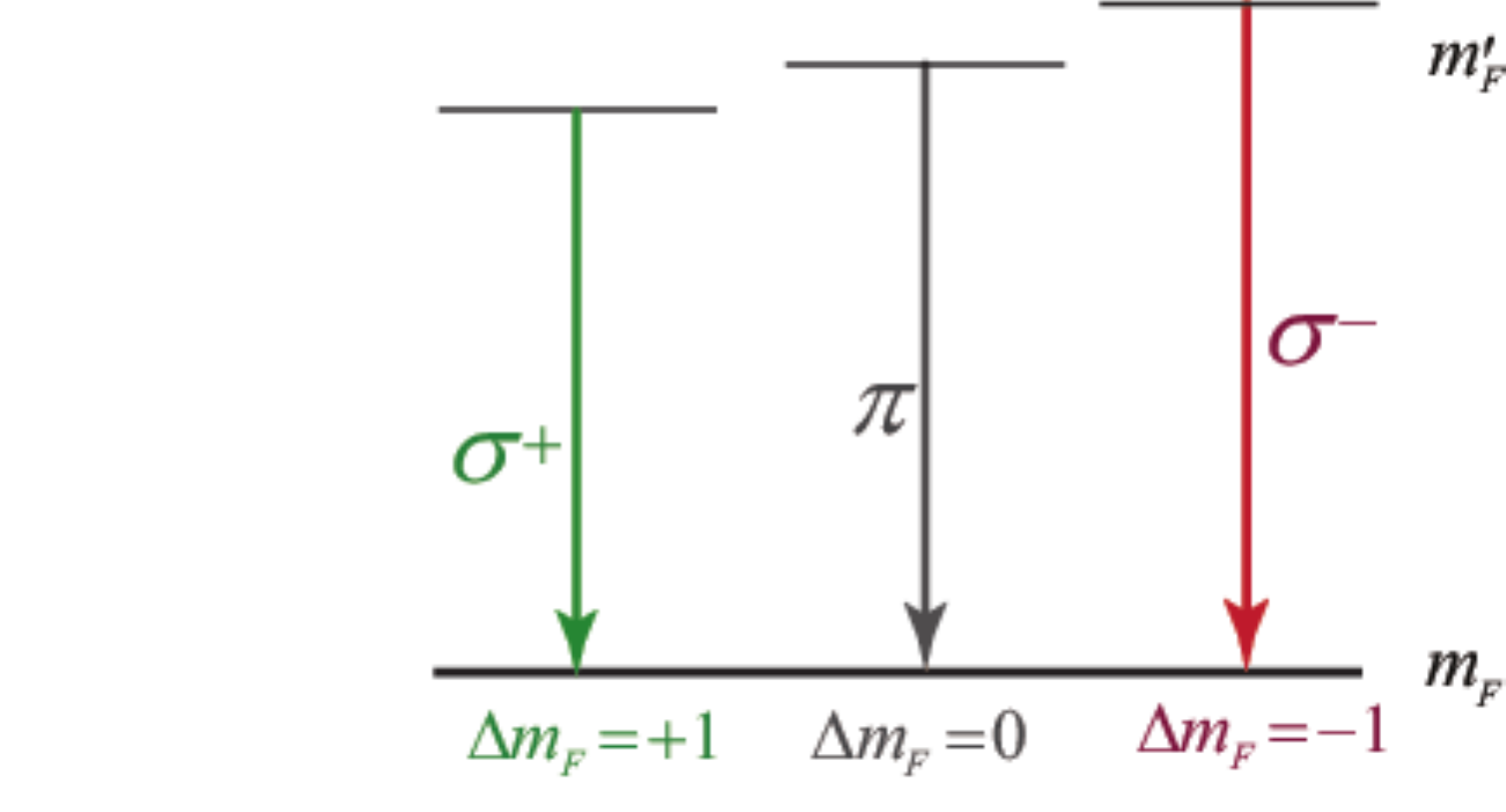}
        \caption{}
        \label{Fig:Zeeman_Transitions}
        \end{subfigure}
        \caption{Schematic of the proposed experiment to study spin photonics in WGMs. The unique proposed effect due to the locked electromagnetic triplet consisting of spin, momentum, and decay. (a) A quantum source with circularly polarized emission ($\sigma^\pm$ transitions) is placed in the vicinity of a spherical resonator. The near-field interaction between the source and TM WGMs of the sphere results in excitation of WGM with only spin polarized, positive OAM along $z$ direction. This unidirectional behaviour is a manifestation of spin-momentum locking in a 3D structure. Spin, linear momentum, and decay are along $\hat{\theta}$, $\hat{\phi}$, and $\hat{r}$, respectively, and form a triplet for the TE and TM modes. (b) General form of Zeeman transitions in a cold atom \cite{sague2007cold} or quantum dot \cite{bayer2002fine}. For $\sigma^\pm$ and $\pi$ transitions, $\Delta m_F = \pm 1$ and $\Delta m_F = 0$, respectively, where $m_F$ is the quantum number pertinent to the total angular momentum of the source (nucleous and electrons). These transitions can be modeled by dipole sources with the electric dipole moment given by Eq. \ref{Eq:Dipole_moment} \cite{mitsch2014quantum}.}
        \label{Fig:Full_Schematic}
\end{figure}

In this work, we present a manifestation of spin-momentum locking with whispering gallery modes of a 3D spherical microresonator. Spin-momentum locking arises naturally through Maxwell's equations for evanescent electromagnetic fields \cite{van2016universal}, resulting in a well-defined triplet for the spin, momentum, and decay vectors (shown in Figure \ref{Fig:Schematic}). Due to the strong field confinement of whispering gallery modes, we show that these modes follow similar spin-momentum locking behavior. We should note that
while the photonic spin of guided modes can be probed by optical force measurements \cite{kalhor2016universal,alizadeh2015transverse,wang2014lateral,hayat2015lateral}, probing the electromagnetic spin of an emitter faces complexities due to the interaction between the source and the probing system in the near-field limit. In fact, a proper definition of photonic spin in the presence of sources remains an open question for this reason \cite{barnett2016natures}.

In this paper, we show that the whispering gallery modes of a spherical resonator form an excellent platform for studying the interaction of spin-polarized quantum radiation sources and the electromagnetic spin of confined modes. In particular, we show that the spin of an emitter effectively couples to the local spin density of whispering gallery modes and ultimately gives rise to the unidirectional propagation of orbital angular momentum modes inside the spherical resonator. The origin of the electromagnetic spin of a quantum emitter is the atomic $\sigma^\pm$ transitions (shown in Figure \ref{Fig:Zeeman_Transitions}) which can be modeled by a circularly polarized dipole.

Using a numerically exact 3D Dyadic Green function approach, we show it is possible to selectively excite particular TE and TM modes with specific radial ($n_r$) and total orbital angular momentum ($l$) numbers. Moreover, by coupling to the spin of TM modes, we show it is possible to induce unidirectional coupling between the Zeeman transitions of an atom \cite{sague2007cold} or a quantum  dot \cite{bayer2002fine} and the whispering gallery modes with either positive or negative orbital angular momentum. Similar observations have been made for 2D WGMs in microdisk resonators \cite{redding2012local,aoki2006observation} as well as 3D WGMs of spherical resonators \cite{shomroni2014all}. There is, however, to the best of our knowledge,
a theoretical gap in the studies of spin properties of whispering gallery modes in a 3D spherical resonator due to the added complexity \cite{setala2002degree}.
Our results should be experimentally observable by methods using tapered fiber coupling \cite{shomroni2014all,vahala2003optical} where whispering gallery modes with positive (negative) orbital angular momentum propagate only along the positive (negative) direction inside the fiber.

\section{Photonic Spin in Spherical Whispering Gallery Modes}
The modes of a spherical resonator are found by solving Maxwell's equations using appropriate boundary conditions in the spherical coordinate representation \cite{jackson2012classical,chiasera2010spherical}. Each mode is labeled by three eigennumbers: $n_r,l,$ and $m$ where $n_r = 1,2,3,\cdots$ is the radial eigennumber while $l$ and $m = -l , -l+1 , \cdots , +l-1, +l$  denote the orbital angular momentum eigennumbers through the eigenvalue relations $\pmb{L}^2 \psi = l(l+1) \psi$ and $L_z \psi = m \psi$, where $\psi$ is either the electric or magnetic field, $\pmb{L} = \pmb{r}\times \nabla$, and $L_z = - i \frac{\partial}{\partial_\phi}$. These relations indicate that $m$ is the projection of OAM along the $z$ axis and modes with positive (negative) $m$ are those that orbit the $z$ axis counter-clockwise (clockwise). For a perfect spherical resonator, the eigenfrequency depends only on $n_r$ and $l$, therefore an emitter with a fixed transition frequency can only selectively couple to $l$ modes but not $m$ modes.
Whispering gallery modes are further distinguished by their polarization, denoted as transverse-electric (TE) modes ($\mathbf{E}(\mathbf{r},\omega)\cdot \mathbf{r}=0$) or transverse-magnetic (TM) modes ($\mathbf{H}(\mathbf{r},\omega)\cdot \mathbf{r}=0$).
For the rest of the paper, we will distinguish these two types of modes using the labels TE$_{n_r,l,m}$ and TM$_{n_r,l,m}$.

Orbital angular momentum and spin are distinctly different properties of the fields.
While orbital angular momentum is a global property, photonic spin is a local property related to the rotational symmetry of the spin-1 electromagnetic vector field \cite{barnett2016natures}. This difference is revealed by observing how spin-polarized sources interact with the whispering gallery modes locally. As one might expect, placing the spin-polarized source in the vicinity of spherical resonator should generate modes with positive OAM. However, as shown in the next section, the exact opposite happens. Spin-polarized source excites WGMs with an OAM that is anti-parallel to the spin of source. This can only be explained by the interplay between the spin of the source and the local spin of the WGMs resulting in the generation of scattered fields that have their OAM anti-parallel to the spin of the source. This observation shows that the spin-polarized source couples to the local spin of the WGMs and not their OAM.
\begin{figure}
    \centering
    \includegraphics[width=0.6\textwidth]{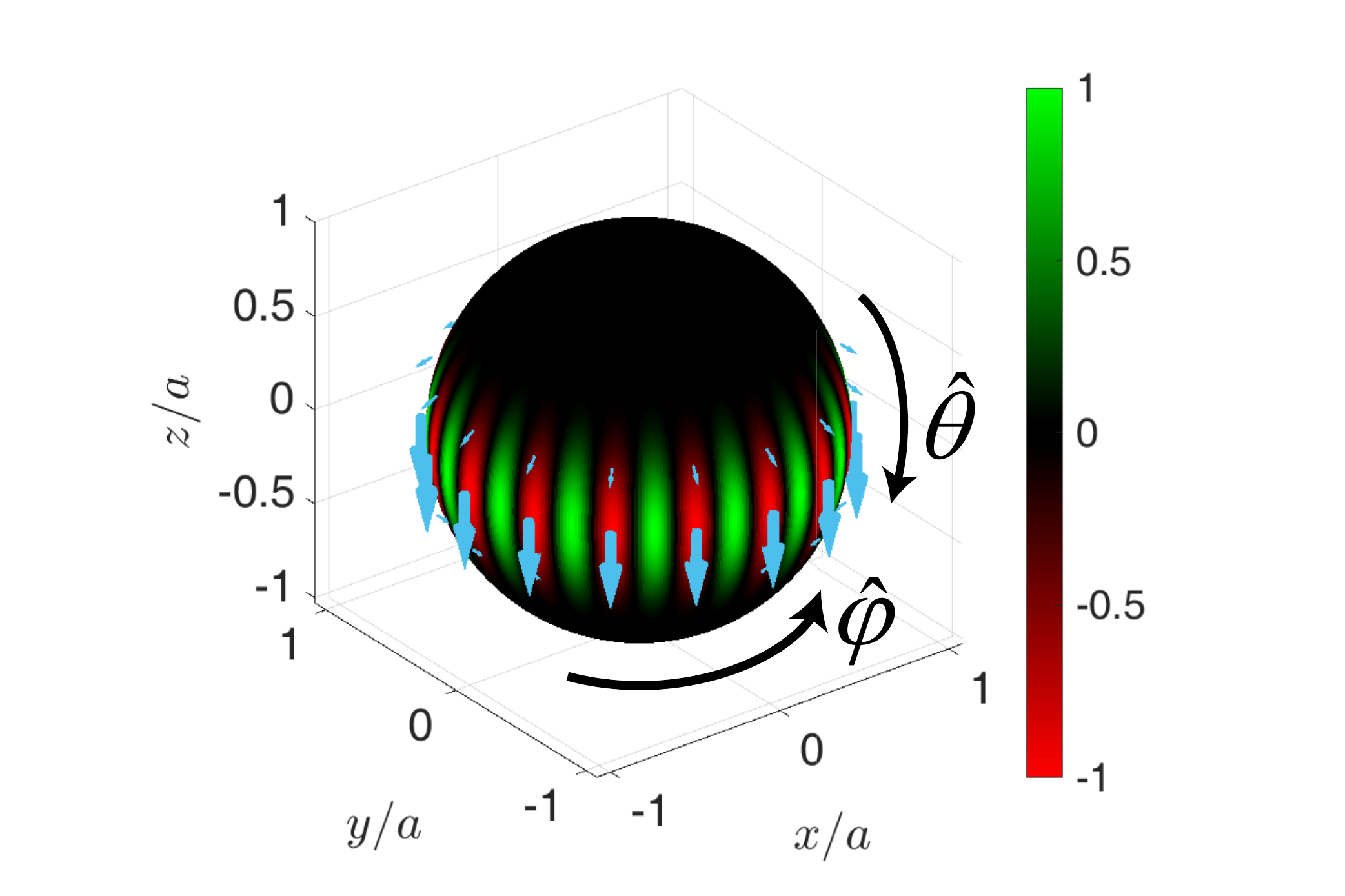}
    \caption{Electromagnetic spin in TE and TM whispering galley modes. The color plot shows the field intensity of $H_r$ ($E_r$) component of the TE (TM) mode for $l=16$ and $m=16$ on the surface of the resonator. The blue arrows show the direction of spin on the surface of the sphere. Modes with positive $m$, orbit the $z$ axis counter-clockwise ($+\hat{\phi}$) while those with negative $m$ orbit the $z$ axis clock-wise ($-\hat{\phi}$). With linear momentum along $+\hat{\phi}$, Momentum, decay, and spin form a triplet. Spin direction follows the spin-momentum locking property for both TE and TM modes. This means that by changing the direction of OAM (changing the sign of $m$), the direction of the spin (blue arrows) reverses for both TE and TM modes. This behaviour inspires  unidirectional coupling of a circularly polarized dipole to the WGMs.}
    \label{Fig:Spin_WGMs}
\end{figure}

The expression for the local spin density of the electromagnetic field in source-free regions is given by $\pmb{S}(\pmb{r},\omega) = \frac{1}{4\omega}  \text{Im}\{\epsilon_0 \pmb{E}^*(\pmb{r},\omega)\times\pmb{E}(\pmb{r},\omega)  + \mu_0 \pmb{H}^*(\pmb{r},\omega)\times\pmb{H}(\pmb{r},\omega)\}$ \cite{barnett1994orbital,berry1998paraxial,picardi2018janus,picardi2018angular}. From this expression, we see that a circularly polarized plane wave propagating along the $z$-direction in free-space, $\pmb{E}(\mathbf{r},\omega) = E_0 (\hat{x}+ i \hat{y})e^{ikz}e^{-i\omega t}$, has an electromagnetic spin pointing along the $z$-direction. For the rest of the paper, we will drop the arguments $(\pmb{r},\omega)$ for notational simplicity. Using this expression, we can calculate the spatial distribution of the photonic spin density for whispering gallery modes. Figure \ref{Fig:Spin_WGMs} shows the field distribution for the TE and TM modes (color plot) as well as their respective electromagnetic spin (blue arrows) on the surface of the sphere with radius $a$. The plots correspond to the TE$_{1,16,16}$ and TM$_{1,16,16}$ modes for which $\lambda_{TE} = 0.54 a$ and $\lambda_{TM} = 0.52 a$. In particular, the spin can be written as:
\begin{equation}\label{Eq:Spin_TM_Simple}
    \pmb{s} = {\pmb{p}}\times {\pmb{\gamma}}
\end{equation}
where $\pmb{s}$ , $\pmb{p}$, and $\pmb{\gamma}$ denote the unit vectors pointing along the spin, the linear momentum, and the decay directions respectively \cite{kalhor2016universal,van2016universal,pendharker2018spin}, thereby forming a right-hand rule triplet. Note that $\pmb{p}$ and $\pmb{\gamma}$ are defined as the real and imaginary part of the Poyting vector, respectively \cite{picardi2018janus,van2018dirac}.

As shown in Fig. \ref{Fig:Spin_WGMs}, the spin of both TE and TM modes (blue arrows) are dominated by the $\hat{\theta}$ component. Explicitly, the dominant electromagnetic spin components $S_{lm, ~\theta}^{TM}$ and  $S_{lm, ~\theta}^{TE}$ can be written as:
\begin{equation}\label{Eq:Spin_of_WGMs}
        S_{lm, ~\theta}^{TM} = S_{lm, ~\theta}^{TE} = - m \frac{\mu_0}{2\omega}\frac{l(l+1)}{|k_1|^2 a^2 } g(\theta) \left[ \mathcal{R} \left\{k_1 a j^*_l(k_1 a) j_{l+1}(k_1 a) \right\} - (l+1) |j_l(k_1 a)|^2\right],
\end{equation}
with $k_1 = \frac{\omega\sqrt{\epsilon_r}}{c}$ being the propagation constant inside the sphere, $\epsilon_r = 3$ is the dielectric permittivity of the sphere, $\omega$ the angular eigenfrequnecy of the TE or TM mode, $\mu_0$ the vacuum permeability, $j_l(ka)$ the spherical Bessel function of the first kind and order $l$, and $g(\theta)$ a real function of $\theta$. $\mathcal{R}\{\}$ takes the real part of its argument. These expressions are derived for fields on the surface of the sphere. We emphasize that the electromagnetic spin, $S$, is linearly dependent on the azimuthal orbital angular momentum, $m$. This result indicates that the direction of the electromagnetic spin is locked to the direction of $z$-projected orbital angular momentum. In other words, changing the sign of $m$ flips the sign of the spin for both TE and TM modes.

These solutions are found under the assumption that the solutions outside the sphere are decaying. Changing the outside solutions to growing solutions, instead, changes the sign of the expression inside the brackets in Eq.\ref{Eq:Spin_of_WGMs}. This means that under the change of direction in the decay vector, the spin for both TE and TM modes flips sign. Together with the linear dependence on $m$, these observations show the spin-momentum locking property as shown in \cite{van2016universal}, and also the fact that spin, momentum, and decay form a triplet. These properties are manifestations of spin-orbit coupling where the change in the OAM results in a change in the spin of WGMs. These previously overlooked properties of WGMs have important implications which we will discuss in the next section. Note that these properties are valid for arbitrary-sized spherical resonators.

\section{Near-field spin interaction}
We aim to investigate the near-field interaction of Zeeman transitions of a quantum source with the WGMs of a spherical resonator. For such interactions we focus on the $\sigma^\pm$ transitions observed in a cold atom \cite{sague2007cold} or quantum dots \cite{bayer2002fine}. Solutions of the Green function for a source outside of a sphere can be written in terms of WGMs with different $l$ and $m$ as \cite{tai1994dyadic,li1994electromagnetic},
\begin{subequations}\label{Eq:Greens_Functions}
    \begin{equation}
          \overline{\mathbf{G}}_e(\mathbf{r},\mathbf{r'}) = \overline{\mathbf{G}}_{0e}(\mathbf{r,r'}) + \overline{\mathbf{G}}_{es}(\mathbf{r,r'}),
    \end{equation}
    \begin{equation}
       \overline{\mathbf{G}}_{0e}(\mathbf{r,r'}) =  \frac{\hat{r}\hat{r}}{k_0^2}\delta(r-r') + \frac{ik_0}{4\pi} \sum_{l=0}^{\infty} \sum_{m=0}^{l} C_{lm}
        \left\{\begin{array}{cc}
        \mathbf{M}_{lm}^{(1)}(k_0)\mathbf{M}'_{lm}(k_0) + \mathbf{N}_{lm}^{(1)}(k_0)\mathbf{N}'_{lm}(k_0) & r \geq r' \\
        \mathbf{M}_{lm}(k_0)\mathbf{M}^{'(1)}_{lm}(k_0) + \mathbf{N}_{lm}(k_0)\mathbf{N}^{'(1)}_{lm}(k_0) & r \leq r'
      \end{array}\right.,
    \end{equation}
    \begin{equation}\label{Eq:Green - Scattered - Outside}
        \overline{\mathbf{G}}_{es}^{(11)}(\mathbf{r,r'}) = \frac{ik_0}{4\pi} \sum_{l=0}^{\infty}\sum_{m=0}^{l} C_{lm} \left[\mathcal{B}_M \mathbf{M}_{lm}^{(1)}(k_0)\mathbf{M}_{lm}^{'(1)}(k_0) + \mathcal{B}_N \mathbf{N}_{lm}^{(1)}(k_0)\mathbf{N}_{lm}^{'(1)}(k_0)\right],
    \end{equation}
    \begin{equation}\label{Eq:Green - Scattered - Inside}
      \overline{\mathbf{G}}_{es}^{(21)}(\mathbf{r,r'}) = \frac{ik_0}{4\pi} \sum_{l=0}^{\infty}\sum_{m=0}^{l}C_{lm}\left[\mathcal{D}_M \mathbf{M}_{lm}(k_1)\mathbf{M}_{lm}^{'(1)}(k_0) + \mathcal{D}_N \mathbf{N}_{lm}(k_1)\mathbf{N}_{lm}^{'(1)}(k_0)\right].
    \end{equation}
\end{subequations}
where the subscript $e$ indicates that these are the Green's functions for the electric field, while the subscripts $0$ and $s$ refer to the homogeneous and scattered solutions, respectively. The functions $\mathbf{M}_{lm}$ and $\mathbf{N}_{lm}$ are the two transverse solutions of Maxwell's equations \cite{li1994electromagnetic,tai1994dyadic}. The superscript $(1)$ refers to the solutions with the spherical Hankel functions of the first kind, while no superscript implies solutions with spherical Bessel functions of the first kind. Also, the unprimed and primed solutions show the dependence on the location of the observation point ($\mathbf{r}$) and the location of the source ($\mathbf{r'}$), respectively. The superscripts $(11)$ and $(21)$ indicate the scattered solutions outside and inside the sphere, respectively. $C_{lm}$'s are some constants, $k_0$ and $k_1$ propagation constants outside and inside the sphere, respectively, and $\mathcal{B}_M, \mathcal{B}_N, \mathcal{D_M},$ and $\mathcal{D}_N$ are the coefficients found by applying the boundary conditions \cite{li1994electromagnetic,tai1994dyadic}. Note that these solutions are the summation of the modes with different OAM quantum numbers $l$ and $m$. Also, since $\mathbf{M}$ and $\mathbf{N}$ are the solutions without and with the radial field components \cite{tai1994dyadic}, we can consider them as the TE and TM contributions to the WGMs, respectively. We have used these solutions to find the interaction of $\sigma^\pm$ and $\pi$ transitions with the dipole moments \cite{mitsch2014quantum},
\begin{equation}\label{Eq:Dipole_moment}
    \pmb{d}_{\pm} = d_0 \Hat{\pmb{e}}_{\pm} = \frac{d_0}{\sqrt{2}} (\hat{r} \pm i\hat{\phi}), \quad \pmb{d_\pi} = d_0 \hat{x}
\end{equation}
located outside of a lossless spherical resonator with a relative permittivity of $3$ at $r_d = a + 10 n$m, $\theta_d = \pi/2$, and $\phi_d = 0$. Here, we look at the WGMs with $n_r = 1$ and $l = 16$ by setting the wavelength of the source to that of the WGMs for the corresponding $n_r$ and $l$. The radius of the sphere is therefore chosen to be $a = 1177$nm to have the resonance of the desired mode at $\lambda_0 = 610$nm. The sphere is thus located in the near-field region of the source.
    \begin{figure}[t!]
        \centering
        \begin{subfigure}[t]{0.32\textwidth}
        \includegraphics[width=\linewidth]{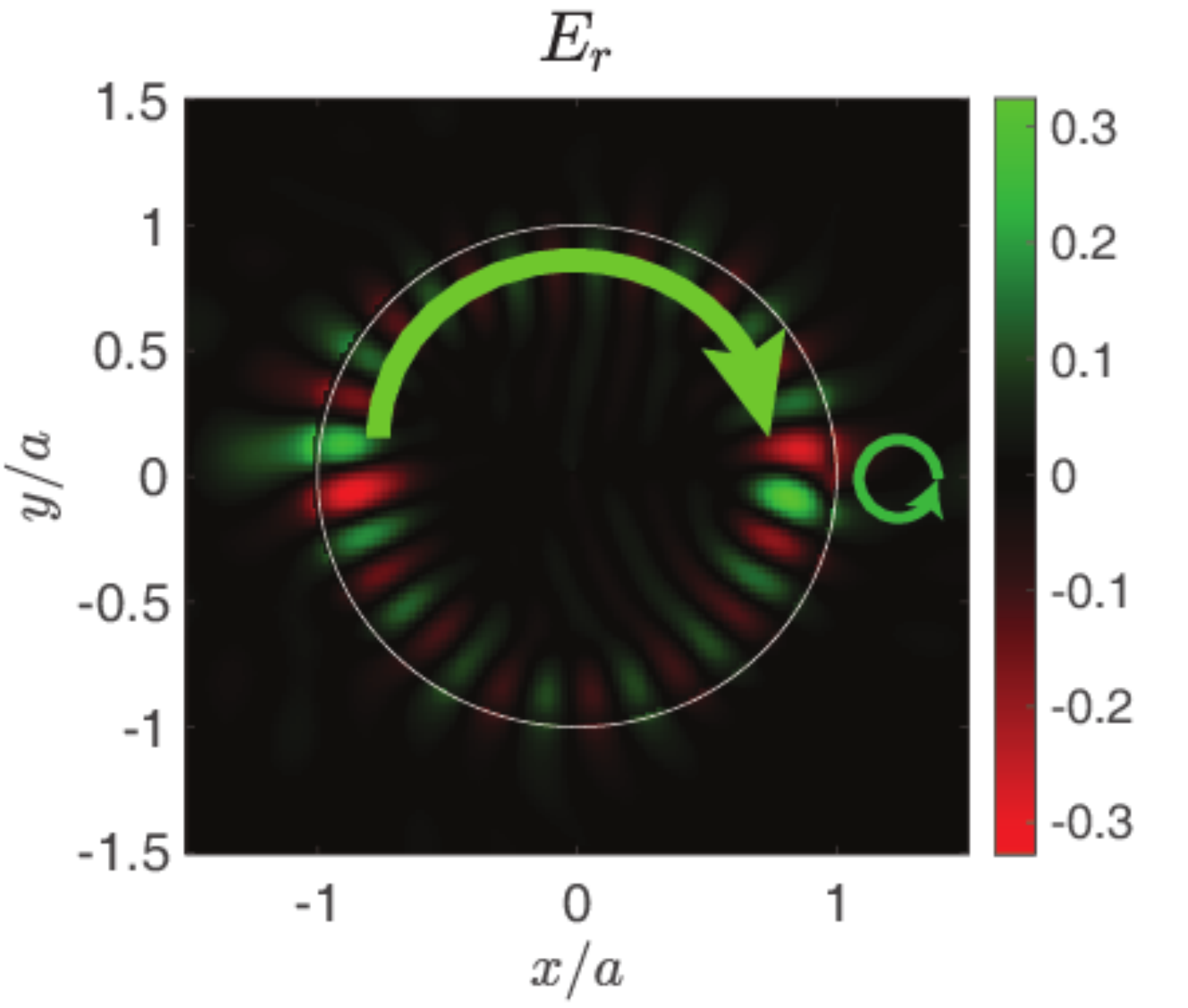}
        \caption{}
        \label{Fig:Perpendicular_dipole_Er}
        \end{subfigure}
        \begin{subfigure}[t]{0.32\textwidth}
        \includegraphics[width=\linewidth]{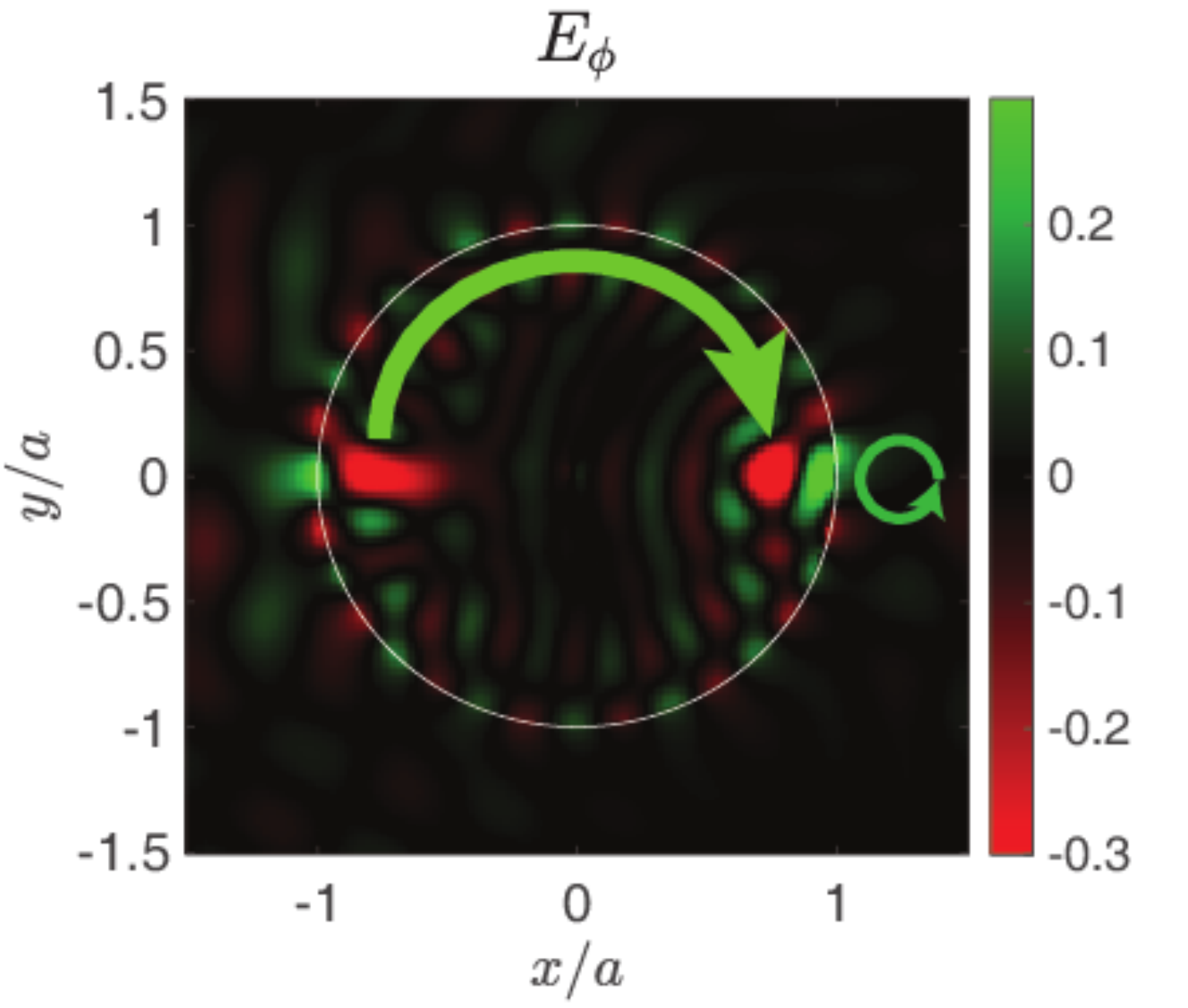}
        \caption{}
        \label{Fig:Perpendicular_dipole_Eq}
        \end{subfigure}
        \begin{subfigure}[t]{0.32\textwidth}
        \includegraphics[width=\linewidth]{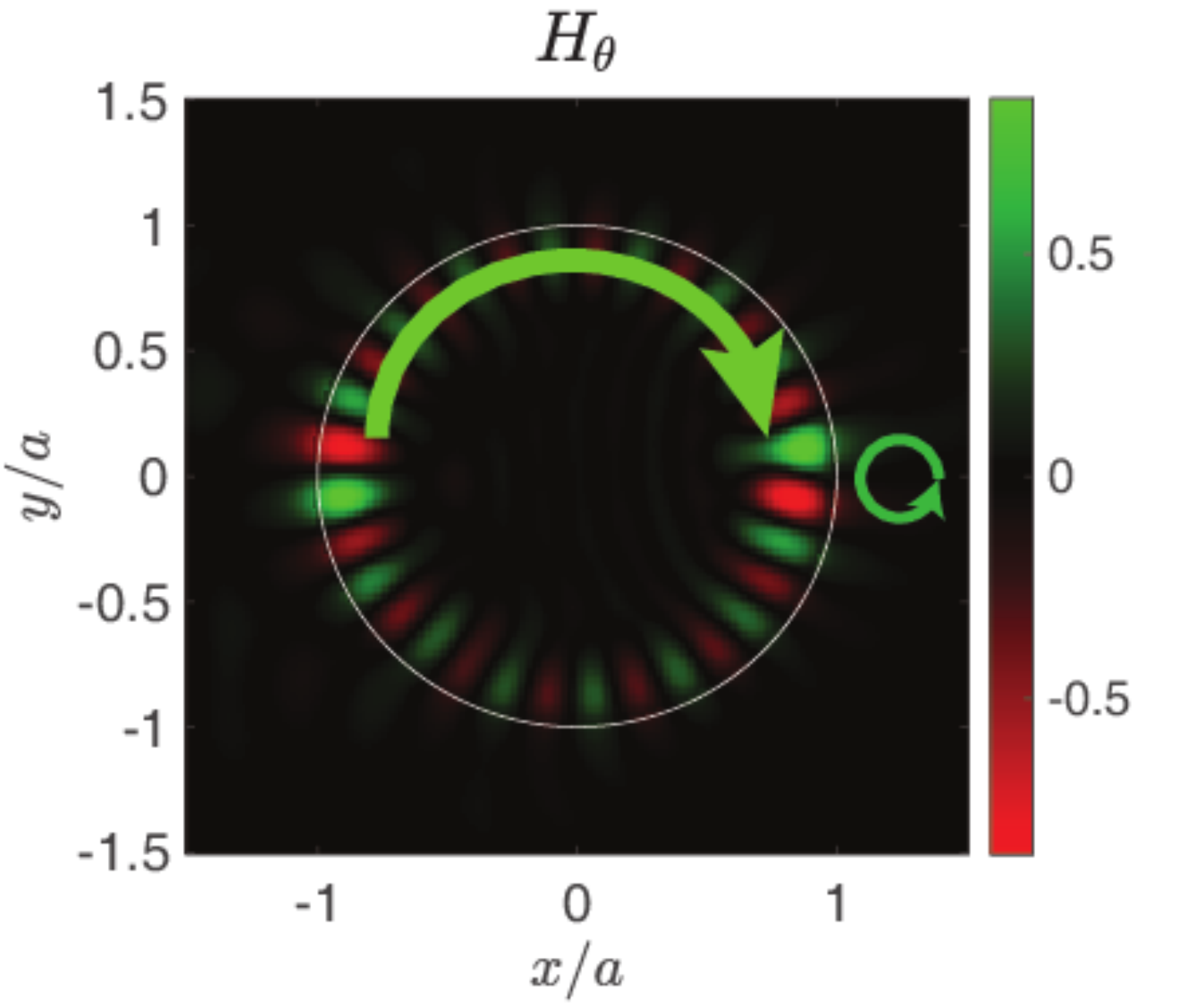}
        \caption{}
        \label{Fig:Perpendicular_dipole_Ho}
        \end{subfigure}
        \caption{Plots of normalized scattered electromagnetic fields due to the right-handed circularly polarized dipole ($\sigma^+$ transition). (a) $E_r$, (b) $E_\phi$, and (c) $H_\theta$ in the $x-y$ plane. All components of the fields orbit along $-\phi$ direction as a result of the circularly polarized dipole located at $x_d = a + 10$nm and $y_d=z_d=0$ with the dipole moment $\pmb{d}_+ = \frac{d_0}{\sqrt{2}} (\hat{x} + i \hat{y})  = d_0 \hat{\pmb{e}}_+$. The circularly polarized dipole couples unidirectionally to the orbit of the fields in the spherical resonator as a result of spin-momentum locking. One important consequence of this is that the photonic spin of the source is opposite to the OAM of the WGMs. Additional videos in supplementary information show the spin-momentum locking (see Visualizations 1 and 2). }
        \label{Fig:Perpendicular_Dipole}
    \end{figure}

Figure \ref{Fig:Perpendicular_Dipole} shows the simulation results for the source with the dipole moment $\pmb{d}_+$ of a $\sigma^+$ transition. Photonic spin of the source in Fig. \ref{Fig:Perpendicular_Dipole} is parallel to the spin of the TE$_{1,16,m>0}$ and TM$_{1,16,m>0}$ modes (Fig. \ref{Fig:Spin_WGMs}). As a result, the dipole excites a mixture of degenerate modes of positive orbital angular momentum along the $z$ direction ($m>0$) and thus gives rise to the unidirectional orbit of the fields inside the sphere. Although spin of the source is parallel to that of both TE and TM modes, only TM modes are excited here. This is due to the fact that the spin of TE mode is primarily from magnetic field while the spin of the TM mode is primarily electric. Having a purely electric spin, the source therefore only couples to the TM mode. This can be equivalently explained by the fact that the TE modes do not have a radial electric field component and therefore they do not couple to the radial component of the dipole moment of the source.

One important observation in Fig.\ref{Fig:Perpendicular_Dipole} is that the photonic spin of the source (pointing out of the plane) is anti-parallel to the orbital angular momentum of the scattered modes inside the sphere (into the plane). This generation of an anti-parallel angular momentum, in the scattered fields, by using a spin-polarized source can only be explained by the fact that the spin of the source is parallel to the local spin of the WGMs (Fig.\ref{Fig:Spin_WGMs}) which results in excitation of modes with anti-parallel OAM. This shows that using a spin-polarized source we can exclusively couple to the photonic spin of the WGMs \cite{barnett2016natures}.

Visualization 1 and Visualization 2 (see online animations) show the clockwise and counter-clockwise rotation of the scattered fields inside the sphere as a result of the circularly polarized dipole located outside of the sphere with the dipole moments of $\pmb{d_+}$ and $\pmb{d_-}$, respectively. This result is an important generalization of spin-momentum locking observed in 1D \cite{barik2018topological} and 2D \cite{petersen2014chiral} problems. In the 3D problem, however, the linear momentum is a result of the orbital angular momentum of the fields.

\begin{figure}[t!]
        \centering
        \begin{subfigure}[t]{0.32\textwidth}
        \includegraphics[width=\linewidth]{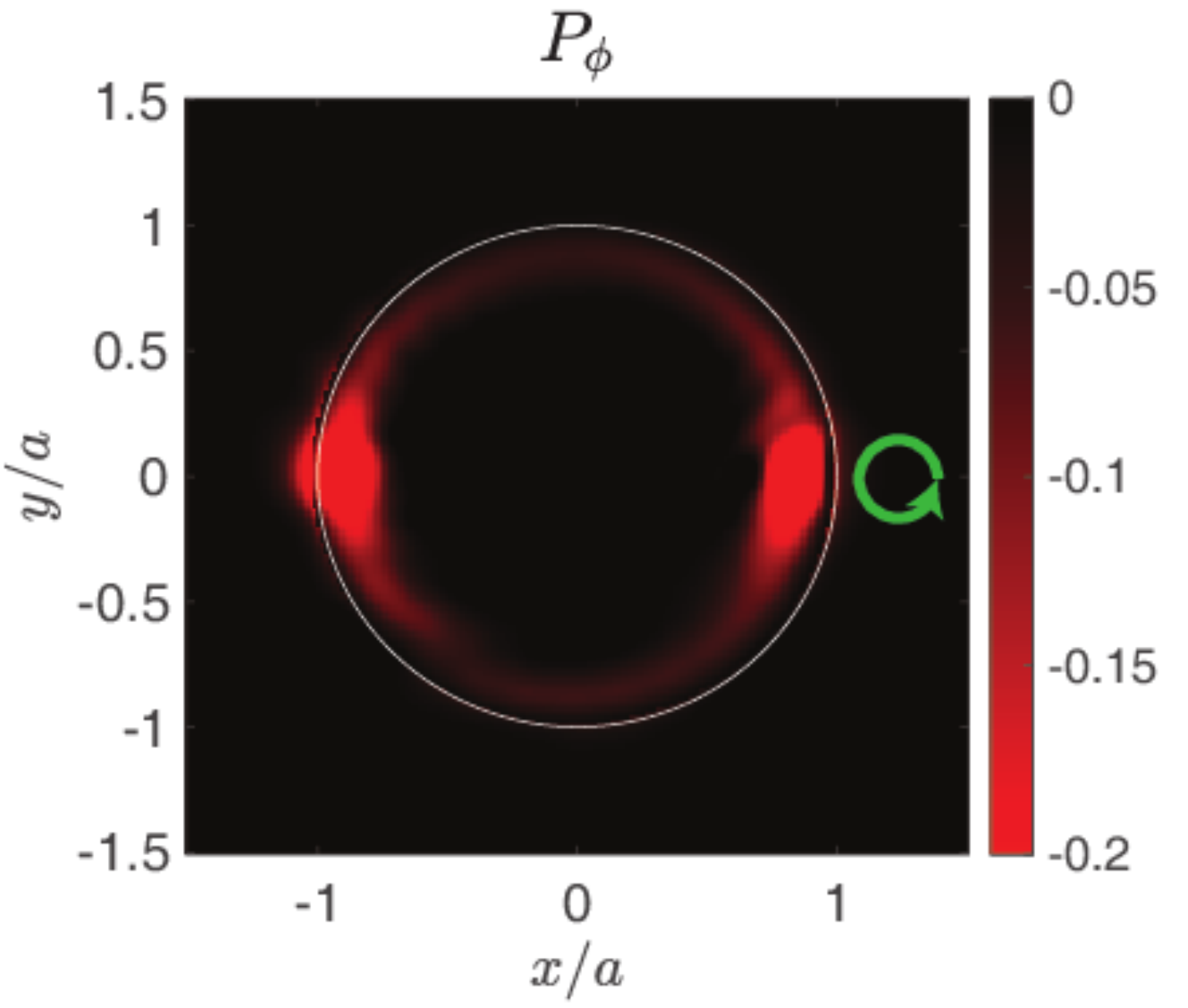}
        \caption{}
        \label{Fig:PoyntingVectors_RH}
        \end{subfigure}
        \begin{subfigure}[t]{0.32\textwidth}
        \includegraphics[width=\linewidth]{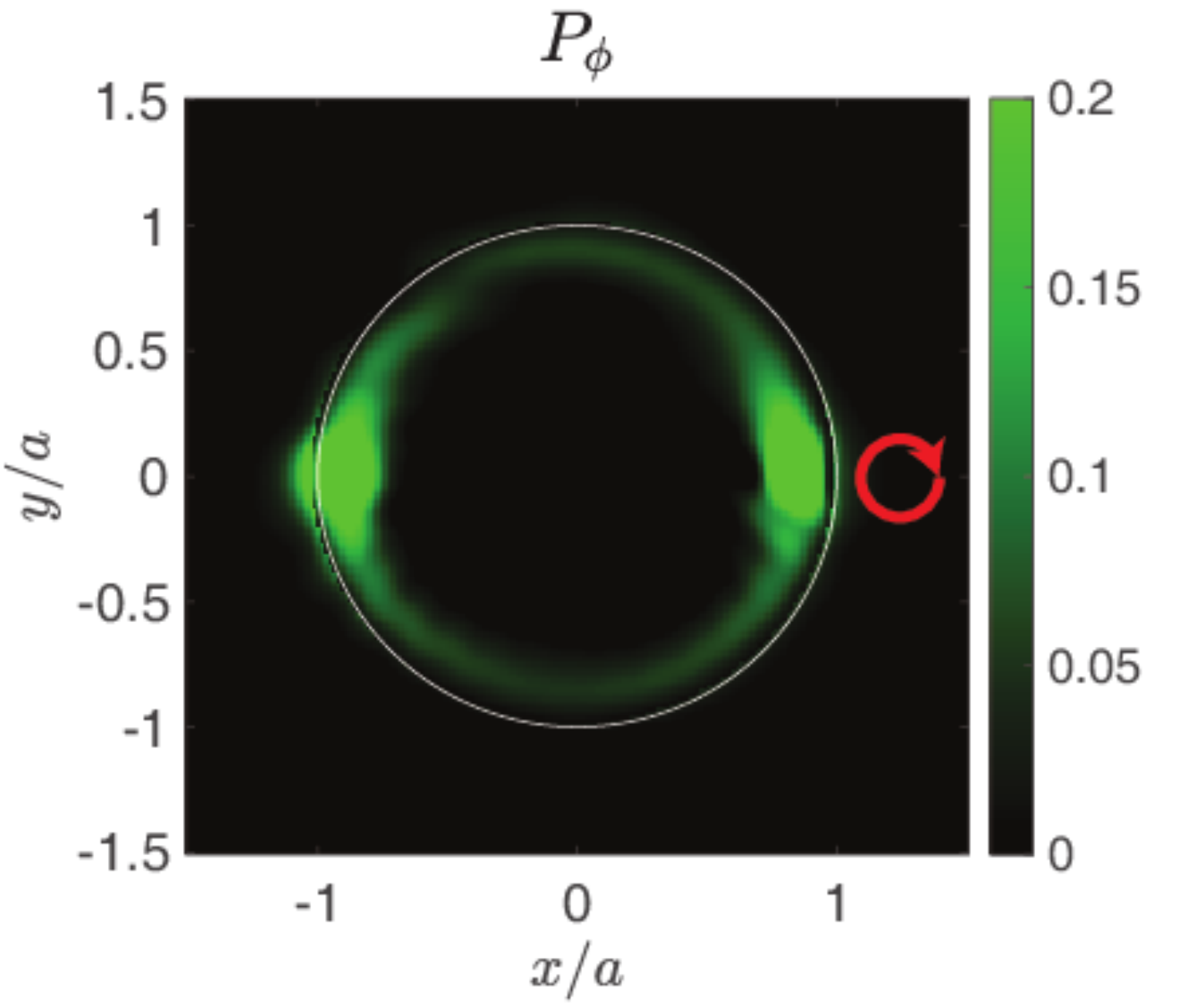}
        \caption{}
        \label{Fig:PoyntingVectors_LH}
        \end{subfigure}
        \begin{subfigure}[t]{0.32\textwidth}
        \includegraphics[width=\linewidth]{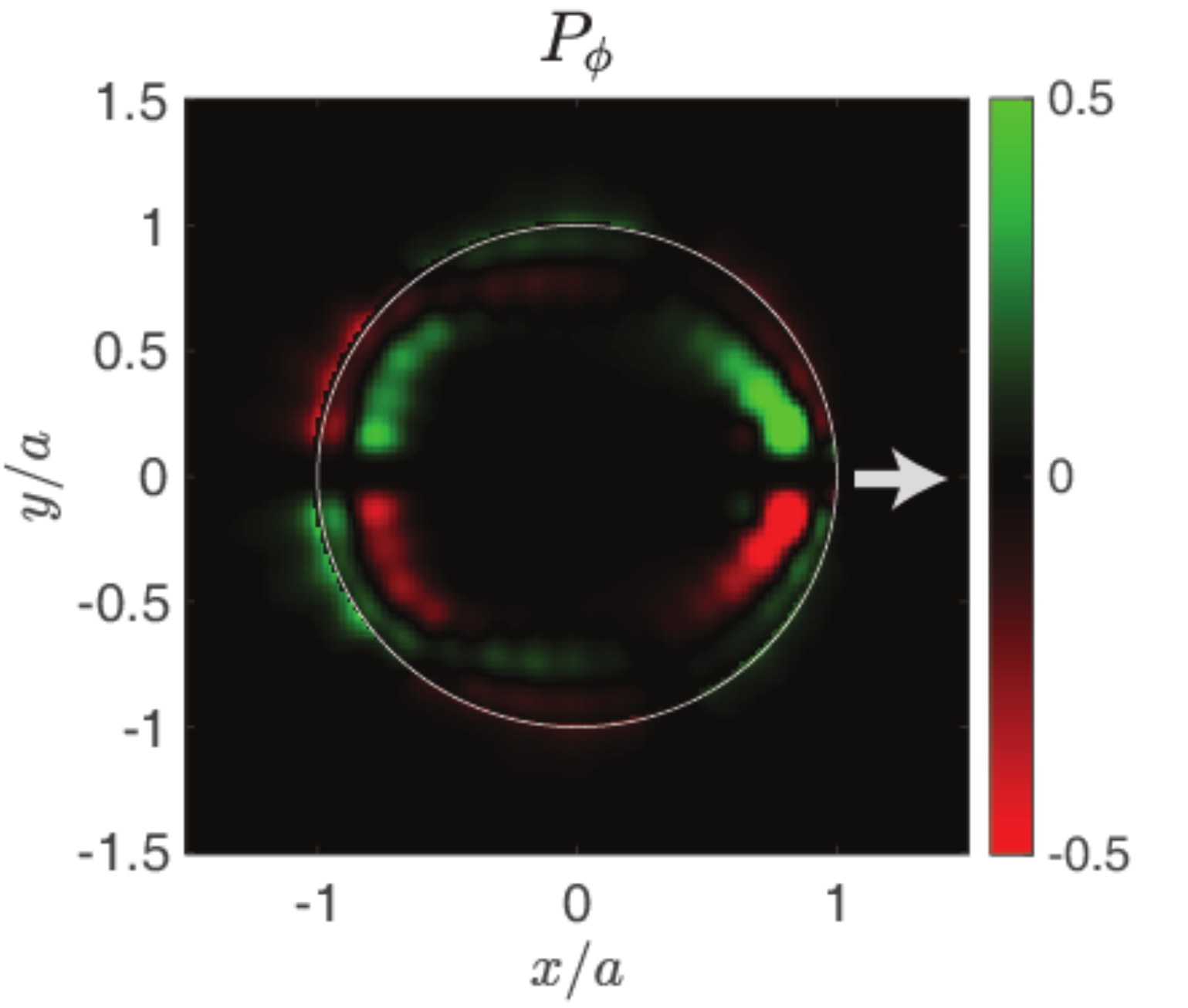}
        \caption{}
        \label{Fig:PoyntingVectors_Lin}
        \end{subfigure}
        \caption{Normalized Poynting vector along $\phi$ direction , $P_\phi$, for the three cases of (a) $\sigma^+$ transitions (RH circularly polarized dipole), (b) $\sigma^-$ transitions (LH circularly polarized dipole), and (c) $\pi$ transitions (linearly polarized dipole along $x$), in the $x-y$ plane for the source located at $x_d = a + 10$nm and $y_d = z_d = 0$, and with the dipole moments given by Eq. \ref{Eq:Dipole_moment}. The negative value of $P_\phi$ in (a) and postive value of $P_\phi$ in (b) indicate that, for the RH and LH circularly polarized dipoles as the source, the WGMs of the spherical resonator orbit clockwise (along $-\hat{\phi}$) and counter-clockwise (along $+\hat{\phi}$), respectively. For the linearly polarized dipole in (c), however, the WGMs inside the sphere are a mixture of clockwise and counter-clockwise fields which eventually cancel out each other to give a net-zero OAM. Therefore, coupling the WGMs to an optical fiber, for instance, on the other side from the source, would result in an equal wave propagation in both directions inside the fiber. However, for a circularly polarized source, the modes would only propagate along one direction inside the fiber, depending on the handedness of source. This figure clearly shows the unidirectional behaviour of spin interaction of the source and WGMs, as a result of the spin-momentum locking.}
        \label{Fig:PoyntingVectors}
    \end{figure}

Figure \ref{Fig:PoyntingVectors} shows the azimuthal Poynting vector, $P_\phi$, inside the sphere, for three cases of right-handed (RH) circularly polarized (Fig.\ref{Fig:PoyntingVectors_RH}), left-handed (LH) circularly polarized (Fig.\ref{Fig:PoyntingVectors_LH}), and linearly polarized (Fig.\ref{Fig:PoyntingVectors_Lin}) dipoles, with the dipole moments given by Eq.\ref{Eq:Dipole_moment}. The dipoles are placed at the same location as that of Fig.\ref{Fig:Perpendicular_Dipole} ($x_d \simeq 1.01a$ and $y_d = z_d = 0$). The unidirectional azimuthal propagation of WGMs inside the sphere is evident as a result of circularly polarized dipole. For the RH dipole (Fig.\ref{Fig:PoyntingVectors_RH}) the Poynting vector is along negative $\hat{\phi}$ (shown as purely red color inside the sphere) meaning that the fields orbit the sphere clockwise, while for the LH dipole (Fig.\ref{Fig:PoyntingVectors_LH}) the Poynting vector is along positive $\hat{\phi}$ (shown as purely green color inside the sphere) meaning that the fields orbit the sphere counter-clockwise. Changing the sense of polarization from RH to LH, changes the sign of azimuthal Poynting vector from negative to positive as seen in Fig.\ref{Fig:PoyntingVectors_RH} and \ref{Fig:PoyntingVectors_LH}. For the linearly polarized dipole in Fig.\ref{Fig:PoyntingVectors_Lin}, however, the fields are a mixture of positively and negatively spinning fields (clockwise and counter-clockwise) which gives a net zero OAM. This result shows that a linearly polarized dipole cannot selectively couple to positive or negative OAM modes, while a circularly polarized dipole can.

To understand this unidirectional behavior further we look at the energy dissipated in the TM WGMs written as \cite{novotny2012principles},

\begin{equation}\label{Eq:H_int}
    \mathcal{W}_{lm}^{TM} = \frac{1}{2} \mathcal{R}\left\{ \pmb{E}_{lm}^{TM}\cdot\pmb{d}_{\pm}^* \right\}.
\end{equation}
where $\pmb{E}_{lm}^{TM}$ is the electric field of the TM$_{1,lm}$ WGM at the location of the source and $\pmb{d_{\pm}}$ is given by Eq.\ref{Eq:Dipole_moment}. The electric TM WGMs fields can be written as \cite{jackson2012classical},
\begin{equation}\label{Eq:E_TM}
    \pmb{E}_{lm}^{TM} = E_{lm,+} \hat{\pmb{e}}_+ + E_{lm,-} \hat{\pmb{e}}_- + E_{lm,\theta}\hat{\theta}
\end{equation}
with
\begin{subequations}\label{Eq:E_TM_Components}
    \begin{equation}
            E_{lm,\pm} =  -\frac{1}{2}\sqrt{\frac{\mu_0}{\epsilon_0}}\left[\frac{l+1}{k_0r_d}f_l(k_0r_d)(l\pm m)\mp f_{l+1}(k_0r_d)\right]Y_{lm}(\theta_d, \phi_d)
    \end{equation}
    \begin{equation}
        E_{lm,\theta} = - \sqrt{\frac{\mu_0}{\epsilon_0}}\left[\frac{l+1}{k_0r_d}f_l(k_0r_d) - f_{l+1}(k_0r_d)\right]\frac{\partial Y_{lm}(\theta_d,\phi_d)}{\partial \theta}
    \end{equation}
\end{subequations}
where $f_l(k_0r_d)$ are the spherical Hankel functions of the first kind evaluated at the location of the dipole, $Y_{lm}(\theta_d,\phi_d)$ spherical harmonics evaluated at the location of the dipole, $k_0$ free space propagation constant, $r_d = a + 10$nm , $\theta_d = \pi/2$, $\phi_d = 0$, and $\Hat{\pmb{e}}_{\pm}$ are given by Eq.\ref{Eq:Dipole_moment}. Note that $E_{lm,+}$ and $E_{lm,-}$ give spin components along $-\Hat{\theta}$ and $+\hat{\theta}$, respectively. We get from Eq.\ref{Eq:E_TM_Components},
\begin{equation}
    \frac{E_{lm,+}}{E_{lm,-}} = \frac{l(l+1)f_l(k_0r_d) - m \left[ k_0r_d f_{l+1}(k_0r_d) - (l+1)f_l(k_0r_d)\right]}{l(l+1)f_l(k_0r_d) + m \left[ k_0r_d f_{l+1}(k_0r_d) - (l+1)f_l(k_0r_d)\right]}
\end{equation}
Note that the terms $\left[ k_0r_d f_{l+1}(k_0r_d) - (l+1)f_l(k_0r_d)\right]$ and $f_l(k_0r_d)$ are always positive for $r_d/a \sim 1 $. Therefore we get,
\begin{equation}
    \begin{split}
        \frac{E_{lm,+}}{E_{lm,-}} \leq 1 , & \quad m \geq0 \\
        \frac{E_{lm,+}}{E_{lm,-}} > 1 , & \quad m < 0
    \end{split}
\end{equation}
This means that according to Eq.\ref{Eq:H_int}, more energy is dissipated into modes with $m<0$ (larger $E_{lm,+}$) for $\pmb{d} = \pmb{d}_+$, while for $\pmb{d} = \pmb{d}_-$, more energy dissipates in modes with $m>0$ (larger $E_{lm,-}$). Since modes with larger $E_{lm,+}$ ($E_{lm,-}$) have their spin along $-\Hat{\theta}$ ($+\Hat{\theta}$), we can say that the spin of $m<0$ ($m>0$) modes aligns with that of the dipole with $\pmb{d}=\pmb{d}_+$
($\pmb{d}=\pmb{d}_-$). Note that although $E_{lm,\theta}$ and $E_{lm,\pm}$ have out-of-phase components, they do not contribute any spin component along $\Hat{r}$ at the location of the source. This means that the photonic spin of the TM WGMs are completely aligning with that of the source.

Using similar expressions and arguments we can show that the dissipated energy into the TE WGMs, as a result of the dipole moment in Eq.\ref{Eq:Dipole_moment}, does not depend on the sign of $m$ because the radial component of the eletric field of the TE WGM is zero. In other words, the TE mode does not show any unidirectional behaviour. Although the photonic spin of the TE mode is parallel to that of the source, the spin of the TE mode is primarily generated by the magnetic field. Since the spin of the source is completely from the electric field (being an electric dipole), a circularly polarized magnetic source should be used to couple to the spin of the TE modes.

Although we have only looked at a particular location of the source, we cannot couple the source to any arbitrary point of the WGMs. This is due to the symmetry of the problem where we essentially choose the $z$ axis (quantization axis) by placing the source in the vicinity of the sphere. Because the total angular momentum of the problem should be conserved, the quantization axis of WGMs (direction of OAM) aligns with the photonic spin of the source. In other words, changing the orientation of the source would also change the quantization axis of the WGMs. For the case when the circularly polarized dipole has no radial component ($\pmb{d}_+ = \frac{d_0}{\sqrt{2}}(\hat{y} + i\hat{z})$ for instance), no spin-momentum locking related phenomenon is observed, as in this case, the spin of the dipole (pointed along $\hat{x}$ direction) is perpendicular to the spin of the TE and TM WGMs.

This unidirectional behavior can be observed by methods such as tapered fiber coupling \cite{vahala2003optical,rosenblum2015cavity} or evanescent coupling \cite{oraevsky2002whispering,shomroni2014all} to the spherical resonator. By coupling the modes of a tapered optical fiber, for instance, to the WGMs of the sphere, unidirectionally orbiting WGMs of the sphere would couple to the optical fiber modes that propagate only in a particular direction. Similar methods to those used in \cite{petersen2014chiral, sague2007cold, shomroni2014all} for a cylindrical problem can be used to trap the source at a particular distance from the sphere and to excite it at the same time. This structure can be an excellent platform to study different forms of spin-spin interaction between electromagnetic fields, atoms, or electrons. Interaction between sources with non-zero electronic spin and the photonic WGMs can be used to understand the near-field spin-spin interaction between the photons and fermions.

\section{Conclusion}
We have presented the theory of spin-momentum locking in 3D whispering gallery modes (WGMs). Our results show that the spin-orbit coupling in WGMs results in modes which form a spin-momentum-decay triplet. This spin-momentum locking property can be observed by coupling the WGMs to the near-fields of $\sigma$ transitions in a cold atom or quantum dot. The results of this paper show that $\sigma^+$ transitions, for instance, only excite TM WGMs with positive OAM. Table \ref{Tab:Summary_of_WGMs} shows the summary of the results of the paper. These results are observable through methods such as tapered fiber coupling or evanescent coupling to the WGMs of the sphere. This structure can be used to study more complex forms of interaction between photonic spin and electronic spin or the interaction of multiple sources with the WGMs.

\begin{table}[!h]
    \centering
    \begin{tabular}{|c|c|c|c|c|}
        \hline
        Whispering Gallery Mode & \multicolumn{2}{c|}{TM$_{lm}$ Mode} & \multicolumn{2}{c|}{TE$_{lm}$ Mode} \\ \hline
        Orbital Angular Momentum & $m>0$ & $m<0$ & $m>0$ & $m<0$ \\ \hline
        Spin & along $+\hat{\theta}$ & along $-\hat{\theta}$ & along $+\hat{\theta}$ & along $-\hat{\theta}$ \\ \hline
        Spin-Momentum-Decay Triplet & \multicolumn{2}{c|}{Yes} & \multicolumn{2}{c|}{Yes} \\ \hline
        Spin-Momentum Locking & \multicolumn{2}{c|}{Yes} & \multicolumn{2}{c|}{Yes} \\ \hline
        Interaction with $\sigma^{\pm}$ Transitions & $\sigma^-$ & $\sigma^+$ & \multicolumn{2}{c|}{No Interaction} \\ \hline
    \end{tabular}
    \caption{Summary of the properties of the WGMs }
    \label{Tab:Summary_of_WGMs}
\end{table}

\section*{Funding}
This work was supported by the Alberta Innovates Technology Future (AITF) scholarship as well as the DARPA Nascent Light-Matter Interactions program.

\bibliographystyle{abbrv}
\bibliography{sample}

\end{document}